\documentclass[conference]{IEEEtran}
\usepackage{cite}
\usepackage{amsmath,amssymb,amsfonts}
\usepackage{algorithmic}
\usepackage{graphicx}
\usepackage{textcomp}
\usepackage{xcolor}
\usepackage{url}
\usepackage{makecell} 
\def\BibTeX{{\rm B\kern-.05em{\sc i\kern-.025em b}\kern-.08em
    T\kern-.1667em\lower.7ex\hbox{E}\kern-.125emX}}
\begin{document}

\title{Optimizing Federated Learning in the Era of LLMs: Message Quantization and Streaming
}
\author{
\IEEEauthorblockN{Ziyue Xu, Zhihong Zhang, Holger R. Roth, Chester Chen, Yan Cheng, and Andrew Feng}
\IEEEauthorblockA{\textit{Nvidia Corp.}, USA}
}

\maketitle

\begin{abstract}
Federated Learning (FL) offers a promising solution for training machine learning models across distributed data sources while preserving data privacy. However, FL faces critical challenges related to communication overhead and local resource constraints, especially in the era of Large Language Models (LLMs) with billions of parameters. The sheer size of these models exacerbates both memory and communication constraints, making efficient transmission and processing essential for practical deployment. NVIDIA FLARE, an open-source SDK for federated learning, addresses these challenges by introducing advanced communication capabilities. Building upon existing solutions for large object streaming, we enhance FL workflows for LLMs through two key techniques: message quantization and container/file streaming. Quantization reduces message size, while streaming enables efficient memory management, improving scalability and integration with existing workflows. These advancements significantly enhance the robustness and efficiency of FL with LLMs, ensuring better performance in real-world federated learning scenarios.
\end{abstract}


\section{Introduction}
Federated Learning (FL) has emerged as a promising approach for training machine learning models across distributed data sources while preserving data privacy. However, FL faces significant challenges related to communication overhead and local resource constraints when balancing model requirements and communication capabilities. Particularly in the current era of Large Language Models (LLMs), the resource requirements to deploy LLMs with billions of parameters pose significant burdens to both local environment and communication capacities - the sheer size of these models exacerbates both communication and memory constraints. Transmitting full model updates in one shot can become infeasible due to bandwidth limitations, while local memory constraints can make processing large models for communication challenging. Addressing these issues requires innovative strategies. 

As LLMs growing bigger, network compression becomes an active research field. Several works have been proposed to compress the model, either by quantization or by pruning, to make the training / evaluation more efficient~\cite{compress, minitron}. Meanwhile, mixed precision training~\cite{mixedprecision} has been widely adopted and is a default for state-of-the-art models like DeepSeek~\cite{deepseekv3}. Low-precision storage, communication, and inference can also be viable solutions~\cite{llama3herdmodels} for efficiency under resource constraints.

\begin{figure*}[bpht]
\centering
\includegraphics[width=\textwidth]{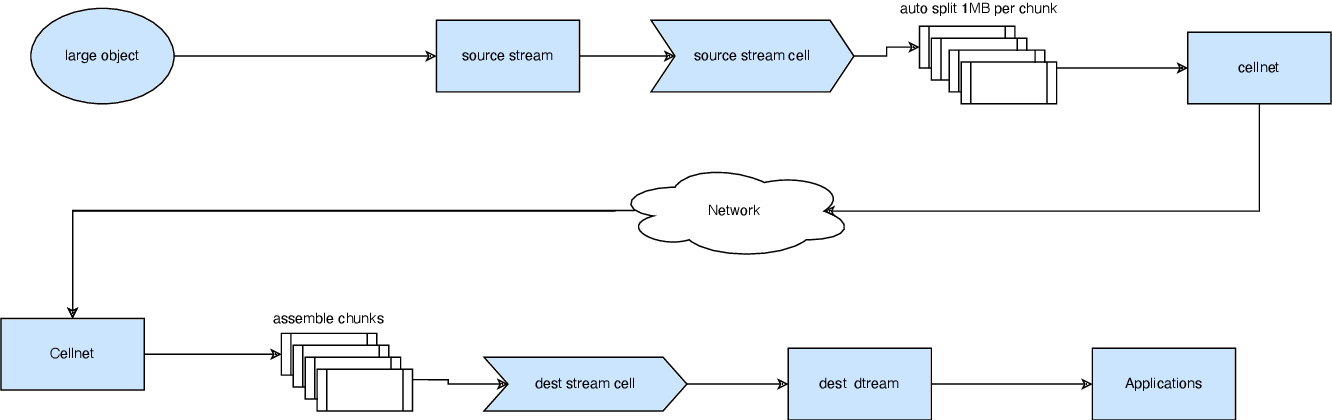}
\caption{NVFlare's Data streaming API.}
\label{fig:1}
\end{figure*}

NVIDIA FLARE (NVFlare)~\cite{nvflare_main}, a domain-agnostic, open-source, and extensible SDK for federated learning, has enhanced the real-world federated learning experience by introducing capabilities to handle communication challenges, including multiple concurrent training jobs, and potential job disruptions due to network interruptions. NVFlare introduced a streaming API~\cite{nvflare_llm} to facilitate the transfer of objects exceeding the 2 GB size limit imposed by gRPC by adding a streaming layer designed to handle large objects by dividing the large model into smaller chunks and stream them to the target. 

To achieve robust and efficient communication for large objects, the data streaming API utilizes the ``Streamable Framed Message'' (SFM)
layer to manage the drivers and connections and sends messages. SFM supports customized drivers without affecting the upper-layer applications. In other words, we can switch between gRPC, TCP, HTTP, etc., and the applications built on top will work without any changes. 

The workflow of data streaming is shown in Figure~\ref{fig:1}, a large model object can be divided into 1 megabyte (MB) chunks and streamed to the target (server or client).%, bringing a complete transformation to the overall system with the introduction of a new streaming layer designed to handle large data transfers. 
Once the message arrives at the target end-point, the object is re-assembled to restore the original message payload.

With the streaming API, we are no longer restricted by the gRPC’s 2 GB size limit, however, as state-of-the-art models growing bigger and bigger(e.g. Llama 3~\cite{llama3herdmodels}), two challenges can still become the bottleneck of an FL workflow with LLMs: the transmission message size under the default fp32 precision, and memory resource requirement to hold the object to be transmitted. The former poses challenge to communication bandwidth and robustness, while the later poses challenge to local hardware resource. 

To enable a more efficient and robust federated pipeline, in NVFlare 2.6.0 release, we introduce two key techniques that facilitate message size reduction and memory-efficient transmission - message quantization and container/file streaming. Specifically:
\begin{enumerate}
\item \textbf{Quantization} and \textbf{de-quantization} for communication implemented as NVFlare’s filters, reducing the message size during transmission.
\item \textbf{Upgraded streaming capabilities} implemented on top of \texttt{ObjectStreamer}, supporting two object types: container and file, with an \texttt{ObjectRetriever} developed for easier integration with existing code.
\end{enumerate}

The main contributions of this work are:
\begin{itemize}
    \item Integration of quantization mechanisms for federated learning with large models like LLMs.
    \item Implementation of efficient streaming capabilities to address real-world challenges under hardware constraints.
    \item Realization of a industry deployment-ready FL pipeline supporting LLM training.
\end{itemize}

\section{Message Quantization: Reducing Communication Overhead}
\subsection{General FL Workflow of NVFlare}
NVFlare's distributed computation framework is built upon the interaction of \texttt{Controller} and \texttt{Executor}. For a basic client-server architecture, we can put the \texttt{Controller} at the server-side, and \texttt{Executors} at remote clients' side: as shown in Fig.~\ref{fig:3}, the \texttt{Controller}, residing on the federated learning server, orchestrates task execution across participating clients. Conversely, \texttt{Executors}, deployed on individual FL client nodes, execute designated computational tasks defined via the client API. The \texttt{Controller}'s operational logic, encapsulated within its $run()$ method, involves task distribution (with `Task Data') to \texttt{Executors} and subsequent aggregation of the returns (with `Task Result') provided by the local task \texttt{Executors}. 

\subsection{Filter Mechanism}
The above workflow design facilitates the modular integration of data transformation pipelines, such as those implementing homomorphic encryption (HE) or differential privacy (DP), to enforce data security and privacy constraints. The outbound and the inbound messages can be filtered before they are packaged for the actual communication. 

Filters can be applied at four points in a round of federated communication, consisting one from server to clients, and one from clients to server:
\begin{itemize}
\item Before `Task Data' leave server
\item Before clients accept `Task Data'
\item Before `Task Result' leave clients
\item Before server accepts `Task Result'
\end{itemize}

Depending on which of these four filters are enabled, we can design proper mechanisms to reinforce post- and pre-processing on the messages

%Take a one-way HE scheme for example, in which case we would like to protect local model updates from being revealed to the aggregator on the server side. In this case:
%\begin{itemize}
%    \item The out-bounding `Task Result' leaving Clients contains local model updates, HE encryption filter will be applied to encrypt it before sending to Server.
%    \item The in-bounding `Task Result' entering Server to be aggregated contains local model updates in cipher-text, no filter applied to it.
%    \item The out-bounding `Task Data' leaving Server contains global weights - i.e., aggregated local model updates in cipher-text, no filter applied to it.
%    \item The in-bounding `Task Data' entering Clients contains global weights in cipher-text, HE decryption filter will be applied to decrypt it before it is received by Clients.
%\end{itemize}

%In this way, all messages communicated between clients and server are cipher-text, the server performs aggregation over cipher-text, avoiding information leakage, while the clients perform training over plain-text. Here, ``one-way'' means encryption and decryption filters are applied on the clients' side only.

\subsection{Two-way Quantization Workflow}
As illustrated earlier, one of the major bottlenecks in FL can be the exchange of model updates among remote participants and the server, as the size of these messages becomes prohibitively large. At our current default message precision at `fp32', this leads to increased latency and bandwidth consumption. Furthermore, given that recent LLMs are trained with mixed and reduced precision, such default precision can artificially inflate the message size. In this case, message quantization offers a solution by reducing the precision of transmitted updates, thereby compressing the message size.

\begin{figure*}[bpht]
\centering
\includegraphics[width=\textwidth]{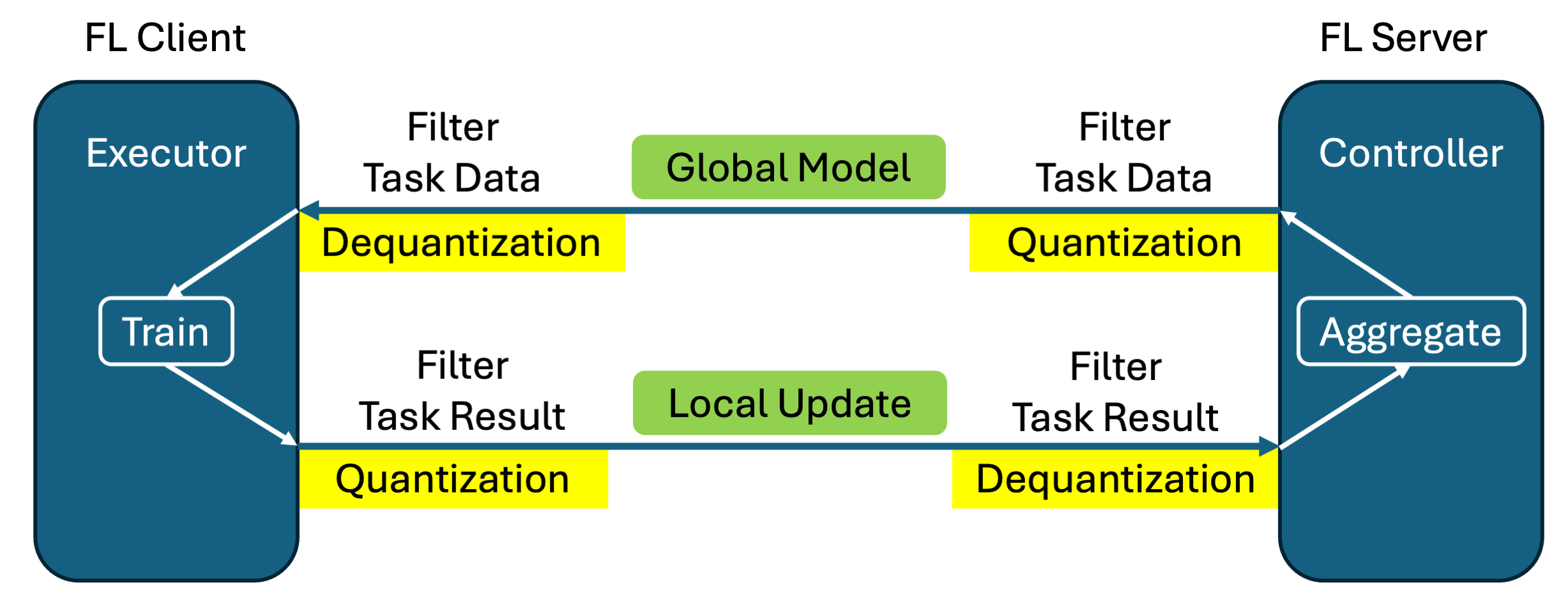}
\caption{Federated Model Training with Message Quantization}
\label{fig:3}
\end{figure*}

As illustrated in Fig.~\ref{fig:3}, we implemented quantization and dequantization with NVFlare's filter mechanism. For message quantization, we will have a two-way filtering scheme applying filters on all four locations:
\begin{itemize}
    \item The out-bounding `Task Result' leaving Clients contains local model updates. A quantization filter will be applied to quantize it to lower precision before sending it to the server for communication efficiency.
    \item The in-bounding `Task Result' entering the server is in quantized state, dequantization filter will be applied to recover it to original precision for aggregation.
    \item The out-bounding `Task Data' leaving the server contains global weights - i.e., aggregated local model updates in original precision, again, the quantization filter will be applied to transform it to lower precision before sending to the clients for communication efficiency.
    \item The in-bounding `Task Data' entering the clients quantized state, and dequantization filter will be applied to recover it to original precision before receiving it by clients for further local training.
\end{itemize}

In this way, all messages communicated between clients and the FL server are in quantized state to save bandwidth and communication cost. At the same time, both server-side aggregation and client-side training are performed with original precision to minimize any loss in precision. In this scheme, ``two-way'' means quantization and dequantization filters are applied on both the server and clients.

There are two benefits of utilizing the filter mechanism: first, no code change will be needed from the model developer - the same training script can be used with and without message quantization with a simple configuration change; second, both training and aggregation will be performed at original precision, rather than quantized data, such that the potential impact message quantization can have over the FL process will be minimized.

\subsection{Quantization Methods}
We use direct cropping and casting to convert fp32 to fp16, and make use of bitsandbytes to perform 8-~\cite{bitsandbytes8} and 4-bit~\cite{bitsandbytes4} quantization. 

\section{Streaming Functionality: Reducing Local Memory Usage}
Another critical challenge in FL is the memory overhead for sending and receiving the messages. Under a one-shot setting, large memory needs to be pre-allocated to hold the entire message for re-assembling the object (although the transmission itself is done by streaming small chunks). Such requirements can be affordable with decent system capabilities and moderate model size, but when considering a LLM with potentially 70B parameters, it can quickly drain the available system memory. 

Considering the fact that even though the entire LLM parameter dictionary can be huge, when breaking down to individual layers - key/parameter pairs, the maximum size of each layer is far smaller, usually around 1 GB. For example, the Llama-3.2-1B model\footnote{\url{https://huggingface.co/meta-llama/Llama-3.2-1B}} has in total 147 layers, including one embed\_token layer, followed by 16 transformer blocks (each with 9 layers), then one norm layer, and finally one lm\_head layer. Each layer has the size listed in Table~\ref{tab:layerwisesize}:

\begin{table}[htbp]
\caption{Layer-wise Sizes of Llama-3.2-1B Model}
\begin{center}
\begin{tabular}{|c|c|}
\hline
\textbf{Layer Name} & \textbf{Layer Size (MB)} \\
\hline
embed\_tokens & 1002.00 \\
layers.(0-15).self\_attn.q\_proj & 16.00 \\
layers.(0-15).self\_attn.k\_proj & 4.00 \\
layers.(0-15).self\_attn.v\_proj & 4.00 \\
layers.(0-15).self\_attn.o\_proj  & 16.00 \\
layers.(0-15).mlp.gate\_proj & 64.00 \\
layers.(0-15).mlp.up\_proj & 64.00 \\
layers.(0-15).mlp.down\_proj & 64.00 \\
layers.(0-15).input\_layernorm & 0.01 \\
layers.(0-15).post\_attention\_layernorm & 0.01\\
norm & 0.01 \\
lm\_head & 1002.00 \\
\hline
\end{tabular}
\label{tab:layerwisesize}
\end{center}
\end{table}

\begin{figure*}[tbph]
\centering
\includegraphics[width=0.75\textwidth]{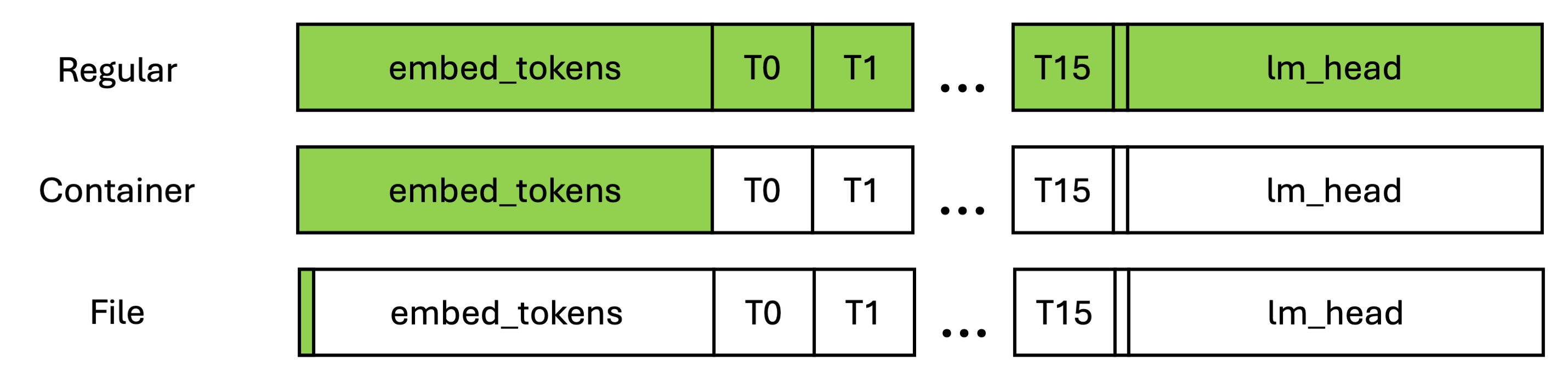}
\caption{Illustration of streaming under different settings, green region indicates the pre-allocated memory needed for communication.}
\label{fig:stream_illu}
\end{figure*}

As shown, although the whole model can be huge, the maximum per-layer size is around 1 GB. Therefore, our upgraded streaming functionality addresses the memory usage challenge by: 
\begin{itemize}
    \item Processing and transmitting the model incrementally, rather than requiring the entire dictionary of gradients to be stored in memory at once. We call it ``container streaming'', which will serialize one item of the parameter dict at a time. So for the above example model with 1 GB maximum per-layer size, the additional memory needed for sending the message is the entire model if sending it as a whole, while \texttt{ContainerStreamer} only needs 1 GB additional memory.
    \item Streaming a file rather than a structured data container. We call it ``file streaming'', which will read the file chunk by chunk and only consumes one chunk of memory. Thus, the additional memory needed is independent of the model size / max item size, and only relies on the the unit streaming size of a file, which can be a very small memory overhead.
\end{itemize}

This process is illustrated in Fig.~\ref{fig:stream_illu}, green boxes mean the maximum local memory that needs to be allocated for the message transmission. As shown, regular transmission needs to allocate memory for the entire model, thus it can be unlimited as the models grow bigger (e.g. 1B to 70B to 400B, etc.). While for \texttt{Container}, the memory is only the same size as the largest layer, which is often bounded by the first \& last layers. Further for \texttt{File}, the memory requirement is independent of the model structure, and is configurable for any file. 

Adapting streaming in FL, we can achieve memory efficiency by breaking down updates into smaller chunks and processing them sequentially. Streaming reduces the peak memory usage, making FL feasible upon resource-model imbalance. Potentially, we can even achieve real-time processing, enabling devices to transmit partial updates while continuing computation, improving responsiveness and reducing idle time. On the receiving side, update strategies can also benefit from adaptive transmission where updates can be sent at varying granularity based on network conditions and client availability.

\section{Experiments and Results}
We used the Llama-3.2-1B model to showcase the functionality of federated SFT, allowing HuggingFace models to be trained and adapted with NVFlare. All other models from HuggingFace can be easily adapted following the same steps. The code is available in NVFlare repo~\footnote{\url{https://github.com/NVIDIA/NVFlare/tree/main/examples/advanced/llm_hf}}. 

\subsection{Quantization Results}
We use the federated Supervised Fine-Tuning (SFT) training scheme to illustrate the quantization impact over federated training. For simplicity, we use a single dataset databricks-dolly-15k\footnote{\url{https://huggingface.co/datasets/databricks/databricks-dolly-15k}}.

We first conduct centralized training with basic SFT training script and single-site federated training. In theory, the two training curves should align, given training randomness, there will be some slight differences.

\begin{figure}[tbph]
\centering
\includegraphics[width=0.5\textwidth]{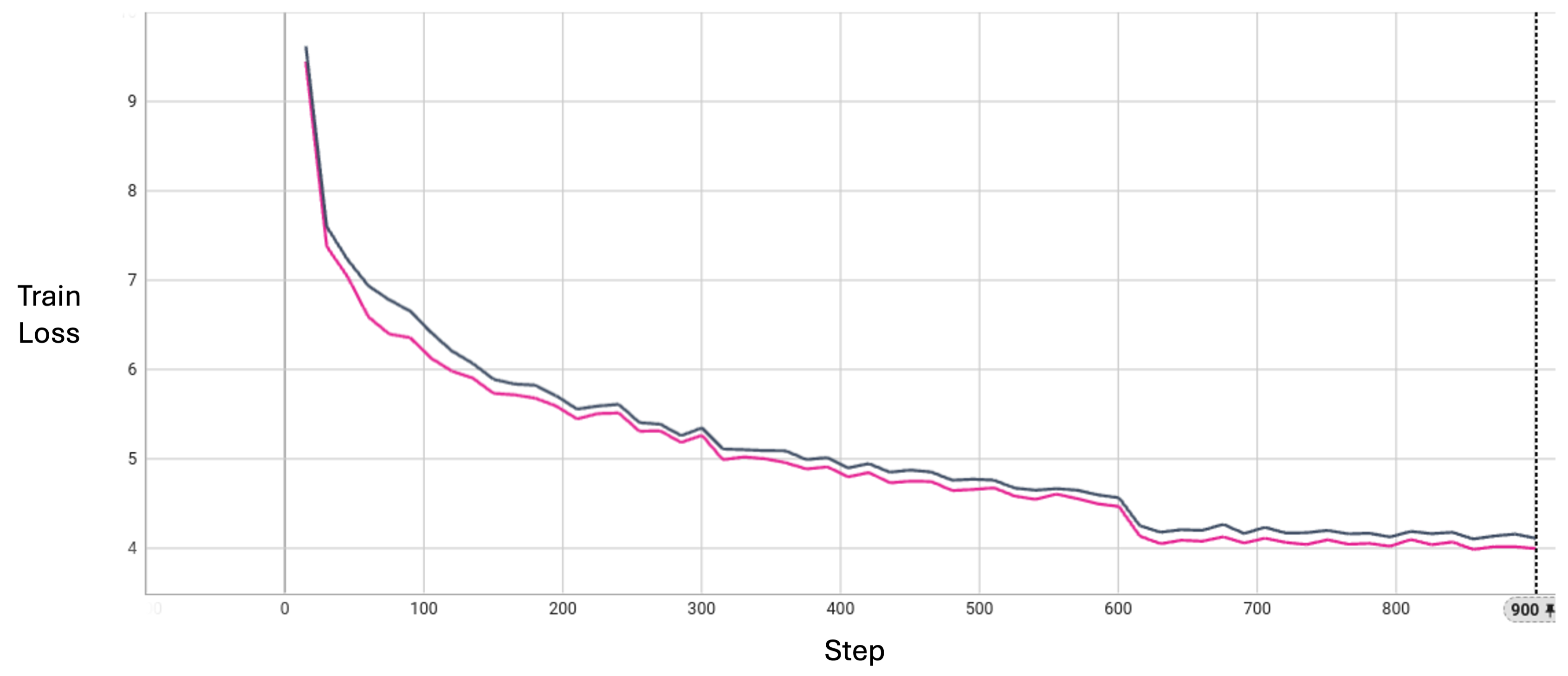}
\caption{Federated SFT comparison: centralized v.s. single-site FL.}
\label{fig:4}
\end{figure}

The resulting training loss curves are shown in Fig.~\ref{fig:4}, as shown, black for centralized results, magenta for FL training. With some training randomness, the two SFT training loss curves align with each other.

Then we conduct the same single-site federated SFT, but this time with various options of message quantization: fp16, blockwise8, float4, and normfloat4. 

\begin{figure*}[btph]
\centering
\includegraphics[width=\textwidth]{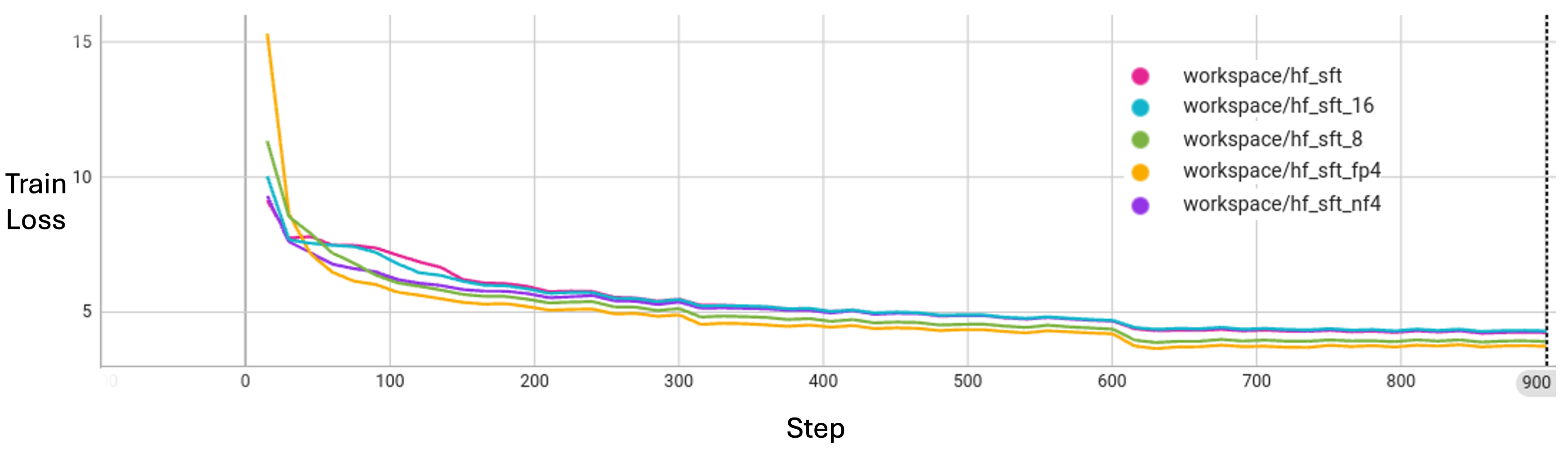}
\caption{Federated SFT comparison: single-site FL with message quantization.}
\label{fig:5}
\end{figure*}

The SFT curves are shown in Figure~\ref{fig:5}, magenta for centralized results, others for FL training with quantization. We can see it achieves similar alignment compared to the centralized result with training randomness as shown in Figure~\ref{fig:4}.

Now let's see the message size reduction via quantization. Table~\ref{tab:1} illustrates the message size in MB for the candidate model under different precisions. As compared with the default fp32, the maximum reduction by 4-bit quantization can reduce the message size to 14\% of the previous default size.

\begin{table}[htbp]
\caption{Message Size under Different Quantization Precisions}
\begin{center}
\begin{tabular}{|c|c|c|c|}
\hline
\textbf{Precision} & \makecell{\textbf{Model} \\ \textbf{Size (MB)}} & \makecell{\textbf{Quantization} \\ \textbf{Meta Size (MB)}} & \makecell{\textbf{fp32 Size} \\ \textbf{Percentage}}\\
\hline
32-bit (fp32) & 5716.26 & 0.00 & 100.00 \% \\
16-bit (fp16, bf16) & 2858.13 & 0.00 & 50.00 \% \\
8-bit & 1429.06 & 1.54 & 25.03 \%\\
4-bit (fp4, nf4) & 714.53 & 89.33 & 14.06 \% \\
\hline
\end{tabular}
\label{tab:1}
\end{center}
\end{table}

By applying message quantization techniques, FL can achieve significant bandwidth savings, and for training LLM with Supervised Fine-Tuning (SFT) in our experiments,  without sacrificing model convergence quality.

\subsection{Memory-efficient Streaming}
For streaming, we use the same 1B model and performed a local simulation of a single global weight communication from the server to a client. The code is available in NVFlare~\footnote{\url{https://github.com/NVIDIA/NVFlare/tree/main/examples/advanced/streaming}}.
Table~\ref{tab:2} illustrates the memory comparisons together with job time information. We record the system memory footprint and compare the peak memory usage of the three settings: regular, container streaming, and file streaming. We can observe that memory usage is significantly reduced when streaming, especially when it comes to file streaming. In contrast, file streaming can take a longer time to finish the job due to file I/O efficiency. 

\begin{table}[htbp]
\caption{Peak Memory Usage under different Streaming Setting}
\begin{center}
\begin{tabular}{|c|c|c|}
\hline
\textbf{Setting} & {Peak Memory (MB)} & {Job Time (s)} \\
\hline
Regular Transmission & 42,427 & 47 \\
Container Streaming & 23,265 & 50 \\
File Streaming & 19,176 & 170 \\
\hline
\end{tabular}
\label{tab:2}
\end{center}
\end{table}

\section{Limitations and Future Work}
The current evaluation, while demonstrating significant message size and peak memory benefits, is constrained by its single-client, single-site setup and the use of a 1B parameter model. This limits our ability to fully characterize the system's performance in a real-world federated learning and task-dependent environment. Specifically, we have not evaluated the convergence stability of repeated quantization/dequantization across multi-client rounds with non-IID data, nor have we explored on the task-specific qualitative metrics. Furthermore, the analysis of end-to-end wall-clock overhead under varying network bandwidths and an evaluation of operational resilience for the streaming mechanism will be helpful for users. Finally, quantization can also impact other essential FL components like Secure Aggregation or Differential Privacy.

Our future work will focus on three key areas to address these limitations. First, extensive multi-client FL evaluations using larger LLMs will be used to validate scalability and convergence with non-IID data, for both LLM training and downstream task finetuning. Second, we plan to deepen the quantization study by investigating per-layer sensitivity and exploring adaptive or error-feedback mechanisms to improve performance at aggressive compression levels. Finally, we will provide a comprehensive systems characterization, including benchmarks for streaming across different chunk sizes and network conditions, and explicitly demonstrate compatibility with other privacy-preserving mechanisms.

\section*{Conclusion}

In this work, we demonstrated how to alleviate communication bottlenecks and memory constraints by integrating message quantization and streaming functionality into FL frameworks. With upgraded capabilities, we make federated learning more efficient and scalable. As these techniques continue to evolve, they will play a crucial role in enabling real-world deployment of FL across diverse environments and applications.
 
Future research can further focus on optimizing finer-grained quantization schemes for various model architectures and developing adaptive streaming mechanisms that dynamically adjust based on network conditions and hardware capabilities.

\bibliographystyle{IEEEtran}
\bibliography{quant_stream}
\end{document}